 \input mssymb.tex

\def\CR{\hbox{{$\cal R$}}}

\def\vect{{\bf t}}
\def\vecu{{\bf u}}\def\vecx{{\bf x}}
\def\veca{{\bf a}}

\def\del{{\partial}}
\def\lform{\hbox{$\sqcup$}\llap{\hbox{$\sqcap$}}}
\def\eps{{\epsilon}}

\def\cocross{{>\!\!\!\triangleleft}}
\def\tens{\mathop{\otimes}}
\def\la{{\triangleright}}

\def\id{{\rm id}}

\def\proof{\goodbreak\noindent{\bf Proof\quad}}

\def\endproof{{\ $\lform$}\bigskip }

\def\und#1{{\underline {#1}}}

\def\o{{}_{(1)}}\def\t{{}_{(2)}}

\def\note#1{}

\def\nquad{{\!\!\!\!\!\!}}
\def\nqquad{\nquad\nquad}
\def\eqn#1#2{\begin{equation}#2\label{#1}\end{equation}}

\def\align#1{\begin{eqnarray*}#1\end{eqnarray*}}
\documentstyle[11pt]{article}
\newtheorem{lemma}{Lemma}[section]
\newtheorem{propos}[lemma]{Proposition}

\newtheorem{theorem}[lemma]{Theorem}

\newtheorem{corol}[lemma]{Corollary}
\newtheorem{defin}[lemma]{Definition}

\begin{document}\baselineskip 22pt

{\ }\hskip 4.7in DAMTP/93-37 %put here preprint number
\vspace{.5in}

\begin{center} {\Large ON THE ADDITION OF QUANTUM MATRICES}
\\ \baselineskip 13pt{\ }
{\ }\\ S. Majid\footnote{SERC Fellow and Fellow of Pembroke College,
Cambridge}\\ {\ }\\
Department of Applied Mathematics \& Theoretical Physics\\ University of
Cambridge, Cambridge CB3 9EW, U.K.
\end{center}

\begin{center}
August 1993\end{center}
\vspace{10pt}
\begin{quote}\baselineskip 13pt
\noindent{\bf ABSTRACT} We introduce an addition law for the usual quantum
matrices $A(R)$ by means of a coaddition $\und\Delta \vect=\vect\tens 1+1\tens
\vect$. It supplements the usual comultiplication $\Delta \vect=\vect\tens
\vect$ and together they obey a codistributivity condition. The coaddition does
not form a usual Hopf algebra but a braided one. The same remarks apply for
rectangular $m\times n$ quantum matrices. As  an application, we construct
left-invariant vector fields on $A(R)$ and other quantum spaces. They close in
the form of a braided Lie algebra. As another application, the wave-functions
in the lattice approximation of Kac-Moody algebras and other lattice fields can
be added and functionally differentiated. \end{quote}
\baselineskip 21.5pt

\section{Introduction}

 In recent years there has been a great deal of interest in quantum matrices.
These algebras are modelled on the co-ordinate functions $t^i{}_j$ on the ring
of matrices $M_n$ say, whose value at matrix $M$ is its component $M^i{}_j$. As
is well-known by now, we keep their abstract algebraic properties but allow the
generators $t^i{}_j$ to be non-commutative. The non-commutativity can
fruitfully be taken in a quadratic form controlled by a matrix $R$ and the
resulting algebras $A(R)$ have nice properties particularly when $R$ obeys the
quantum Yang-Baxter equations (QYBE). Moreover, by adding suitable further
relations, one obtains for example $q$-deformations of the algebras of
functions on the standard compact matric groups\cite{FRT:lie}.

We further recall that the  structure of $A(R)$ is that of a bialgebra with a
coproduct $\Delta:A(R)\to A(R)\tens A(R)$, $\Delta \vect=\vect\tens\vect$
encoding the formal properties of matrix multiplication. This coproduct has
numerous uses, allowing us to tensor product representations and other
applications.

In this paper we develop a further `coaddition' structure $\und\Delta:A(R)\to
A(R)\und\tens A(R)$ corresponding this time to the addition of quantum
matrices. Moreover, it is compatible with  the existing coproduct $\Delta$ in a
way that corresponds to the usual distributivity axiom. Our construction so far
works for $R$ of Hecke type, but this is the most common and applies for
example to the standard $GL_n$ $R$-matrices. In this way we complete the
structure of $A(R)$ to fully model the ring structure -- both addition and
multiplication -- of usual matrices. Such a structure can be called a {\em
quantum-braided ring} for obvious reasons.

In order to formulate this structure we need the notion of a braided group or
braided-Hopf algebra as introduced by the author in \cite{Ma:exa} and already
applied in numerous contexts. For physics, the most important one is perhaps in
\cite{Ma:poi}\cite{Ma:fre}. We will see that the coaddition $\und\Delta$ does
not form a usual quantum group or Hopf algebra, but a braided one.
Nevertheless, it has group-like or vector space-like properties and one can
construct for example braided-differential calculi on $A(R)$ using the
technique of \cite{Ma:fre}. This we do in Section~3. As an application, we
construct left-invariant vector fields on $A(R)$ and see that they close into
some form of braided-Lie algebra.

Let us note in passing that some kind of {\em braided ring} of $2\times 2$
matrices (with both braided comultiplication  and braided coaddition) has
already been found in the work of U. Meyer in his approach to $q$-deformations
of Minkowski space\cite{Mey:new}. This is different from our result but can be
considered as one of our indirect motivations.

We also note that in \cite{MaMar:glu} has been introduced a notion of
rectangular quantum matrices $A(R_1:R_2)$ associated to any pair of
$R$-matrices. We will see in Section~4 that these too can be added via a
coaddition, at least when $R_i$ are of Hecke type. Likewise, one may
differentiate with respect to them. As an application, we construct rotational
vector fields on quantum planes. We also propose a systematic approach to
lattice wave-functions and their functional differentiation.

Finally, in the Section~5 we mention one context in the physics literature
where these results have an immediate consequence. This is to the lattice
approximation of Kac-Moody algebras in \cite{AFS:hid}. In this work there are
quantum-matrix valued `wave-functions' on the discretised line. We will see
that the relations between them introduced in \cite{AFS:hid} have just the
structure of a rectangular quantum matrix and hence can be added pointwise via
our braided coaddition.

\section{Braided-coaddition on quantum matrices}

Here we prove the addition law on $A(R)$ for $R$-matrices of Hecke type. Recall
that a matrix solution of the QYBE $R_{12}R_{13}R_{23}=R_{23}R_{13}R_{12}$ is
Hecke if
\eqn{hecke}{ (PR-q)(PR+q^{-1})=0}
for suitable $q$. Here $P$ denotes the usual permutation matrix. Thus they have
two eigenvalues in their minimal polynomial. We use the standard notations
where the suffices on $R$ etc refer to the position in a matrix tensor product.

We recall also that a braided group (or braided-Hopf
algebra)\cite{Ma:bra}\cite{Ma:exa} is $(B,\und\Delta,\und\eps,\und S,\Psi)$
where the first four are like the coproduct, counit, antipode (when it exists)
of a usual Hopf algebra but $\und\Delta:B\to B\und\tens B$ is an algebra
homomorphism not to the usual tensor product algebra but to the braided tensor
product one. This is defined by a linear map $\Psi:B\tens B\to B\tens B$ which
obeys the braid relations and which is compatible with the other maps in the
sense that the braiding commutes with them (one says that the braiding is
functorial) applied to either input in an obvious way. The braided tensor
product is then \cite{Ma:bra}
\eqn{btens}{ (a\tens b)(c\tens d)=a\Psi(b\tens c)d,\qquad \forall a,b,c,d}
where we apply $\Psi$ and then multiply the result from the left by $a$ and
from the right by $d$ as shown. Thus the notion of a braided group or
braided-Hopf algebra is a generalisation of the notion of super-group or
super-Hopf algebra but with bose-fermi statistics replaced by braid ones.
An introduction for physicists is in \cite{Ma:introp}. The formal mathematical
picture is in \cite{Ma:bg}.

\begin{theorem} Let $R$ be a solution of the QYBE of Hecke type and $A(R)$ the
usual bialgebra as in \cite{FRT:lie}cf\cite{Dri} with a matrix of generators
$\vect=(t^i{}_j)$ and relations $R\vect_1\vect_2=\vect_2\vect_1R$. This forms a
braided-Hopf algebra with
\[ \Psi(\vect_1\tens\vect_2)=R_{21}\vect_2\tens\vect_1R,\quad \und\Delta
\vect=\vect\tens 1+1\tens \vect,\quad \und\eps\vect=0,\quad\und S\vect=-\vect\]
\end{theorem}
\proof We give a direct proof. By definition, $\und\Delta$ extends to products
as an algebra homomorphism to the braided tensor product algebra. This is
consistent because
\align{\und\Delta R\vect_1\vect_2&=& R(\vect_1\tens
1+1\tens\vect_1)(\vect_2\tens 1+1\tens\vect_2)\\
&=&R\vect_1\vect_2\tens 1+1\tens
R\vect_1\vect_2+RR_{21}\vect_2\tens\vect_1R+R\vect_1\tens\vect_2\\
\und\Delta\vect_2\vect_1 R&=& (\vect_2\tens 1+1\tens\vect_2)(\vect_1\tens
1+1\tens\vect_1)R\\
&=& \vect_2\vect_1R\tens 1+1\tens\vect_2\vect_1R+ R\vect_1\tens\vect_2
R_{21}R+\vect_2\tens\vect_1 R}
which are equal because $R_{21}R=1+(q^{-1}-q)PR$ and $RR_{21}=1+(q^{-1}-q)RP$
from (\ref{hecke}).

For a full picture here, we also have to check that $\Psi$ shown here on the
generators extends consistently to products of the generators in such a way as
to be functorial with respect to products. The strategy is like that in
\cite{Ma:exa}\cite{Ma:poi} and reduces to the QYBE for $R$. Explicitly, one has
\align{ \Psi(\vect_1\vect_2\cdots\vect_M\tens\vect_{M+1}\cdots\vect_{M+N})
&&\nqquad=R_{M+1\, M}\cdots R_{M+11}\phantom{\vect_{M+1}\cdots
\vect_{M+N}\tens\vect_1\cdots\vect_M} R_{1M+N}\cdots R_{1M+1}\\
&&\quad\vdots\qquad\qquad\vdots\qquad\vect_{M+1}\cdots
\vect_{M+N}\tens\vect_1\cdots\vect_M\qquad\vdots\qquad\qquad\vdots\\
&&\nquad R_{M+N\, M}\cdots R_{M+N1}\phantom{\vect_{M+1}\cdots
\vect_{M+N}\tens\vect_1\cdots\vect_M}R_{MM+N}\cdots R_{MM+1}}
where the blocks are to be multiplied in the order shown. We will give an
alternative way to deduce this formula later in the section.
\endproof

Another way to deduce the result without verifying directly is to reduce it to
the problem to the construction of braided vector space Hopf algebras in
\cite{Ma:poi}, of which this is an example. To do this we introduce the
notation $t_I={t^{i_0}{}_{i_1}}$ where $I=(i_0,i_1)$, $J=(j_0,j_1)$ etc are
multindices. Then
\eqn{psimulti}{\Psi(t_I\tens t_J)=t_B\tens t_A{\bf R}^A{}_I{}^B{}_J,\quad {\bf
R}^A{}_I{}^B{}_J=R^{j_0}{}_{b_0}{}^{i_0}{}_{a_0}
R^{a_1}{}_{i_1}{}^{b_1}{}_{j_1}.}
\eqn{A(R)multi}{t_It_J=t_Bt_A{\bf R}'{}^A{}_I{}^B{}_J,\quad {\bf
R}'{}^A{}_I{}^B{}_J=R^{-1}{}^{i_0}{}_{a_0}{}^{j_0}{}_{b_0}
R^{a_1}{}_{i_1}{}^{b_1}{}_{j_1}.}
This is now exactly the framework of a covector braided group\cite{Ma:poi} and
one just has to verify that these matrices ${\bf R},{\bf R}'$ obey the
conditions there. These are $(P{\bf R}+1)(P{\bf R'}-1)=0$ which reduces to $R$
Hecke, and mixed QYBE-like relations of the form ${\bf R}_{12}{\bf R}_{13}{\bf
R}'_{23}={\bf R}'_{23}{\bf R}_{13}{\bf R}_{12}$ and ${\bf R}'_{12}{\bf
R}_{13}{\bf R}_{23}={\bf R}_{23}{\bf R}_{13}{\bf R}'_{12}$. One has to write
these out and see that they reduce to the QYBE for $R$ many times. Finally,
these is a matrix condition ${\bf R}'_{21}{\bf R}={\bf R}_{21}{\bf R}'$ which
holds automatically and provides for a braided-antipode.

It is a natural question to ask how this (braided) coaddition structure, which
clearly corresponds to the linear addition of underlying matrices in the
classical case, relates to the usual comultiplication.

\begin{defin} A {\em quantum ring} is a a bialgebra $(B,\Delta,\eps)$ and a
second Hopf algebra structure $(B,\und\Delta,\und\eps,\und S)$ for the same
algebra $B$, which obeys the {\em codistributivity axioms}
\[ (\id\tens\cdot)\circ\Delta_{B\tens
B}\circ\und\Delta=(\und\Delta\tens\id)\circ\Delta\]
\[ (\cdot\tens\id)\circ\Delta_{B\tens
B}\circ\und\Delta=(\id\tens\und\Delta)\circ\Delta\]
where $\Delta_{B\tens B}=(\id\tens\tau\tens\id)(\Delta\tens\Delta)$ is the
usual tensor product coalgebra. Here $\tau$ denotes the usual transposition of
the middle two factors. We call $\Delta$ the {\em comultiplication} and
$\und\Delta$ the {\em coaddition}. If $\und\Delta$ forms a braided Hopf algebra
rather than a usual one, we say that we have a {\em quantum-braided ring}.
\end{defin}

\begin{propos} The braided-Hopf algebra structure on $B=A(R)$ in Theorem~2.1
together with the usual comultiplication $\Delta \vect=\vect\tens \vect$ forms
a quantum-braided ring.
\end{propos}
\proof We have to prove the codistributivity conditions in Definition~2.2.
On the generators they holds trivially. On products $\vect_1\vect_2$ of
generators, we have for the first condition
\align{&&(\id\tens\cdot)\Delta_{B\tens B}(\vect_1\vect_2\tens
1+1\tens\vect_1\vect_2+\vect_1\tens\vect_2+R_{21}\vect_2\tens\vect_1 R)\\
&&=(\id\tens\cdot)\tau_{23}(\vect_1\vect_2\tens\vect_1\vect_2\tens1\tens
1+1\tens1\tens\vect_1\vect_2\tens\vect_1\vect_2+\vect_1\tens\vect_1
\tens\vect_2\tens\vect_2+R_{21}\vect_2\tens\vect_2\tens\vect_1\tens
\vect_1 R)\\
&&=\vect_1\vect_2\tens
1\tens\vect_1\vect_2+1\tens\vect_1\vect_2\tens\vect_1\vect_2+\vect_1\tens
\vect_2\tens\vect_1\vect_2+R_{21}\vect_2\tens\vect_1\tens\vect_2\vect_1R\\
&&=(\vect_1\vect_2\tens 1+1\tens\vect_1\vect_2+\vect_1\tens\vect_2+
R_{21}\vect_2\tens\vect_1R)\tens\vect_1\vect_2\\
&&=(\und\Delta\tens\id)(\vect_1\vect_2\tens\vect_1\vect_2)}
because of the relations in $A(R)$. $\tau_{23}$ denotes transposition in the
middle two factors. One proves the general case in a similar way by induction.
Similarly for codistributivity from the other side. \endproof

For the canonical example, we let
\eqn{Rgl2}{R=\pmatrix{q&0&0&0\cr 0&1&q-q^{-1}&0\cr 0&0&1&0\cr
0&0&0&q}}
the standard R-matrix for the $2\times 2$ quantum matrices $M_q(2)$. This has
generators $\vect=\pmatrix{a&b\cr c&d}$ and the usual relations
\[ qab=ba,\quad ca=acq,\quad qcd=dc,\quad db=dbq,\quad bc=cb,\quad
ad-da=(q\sp {-1}-q)cb.\]
This forms a quantum-braided ring with the usual comultiplication and the
linear braided coaddition on the generators, with the braiding
\align{&\Psi(a\tens a)=q^2 a\tens a &\Psi(a\tens b)=q b\tens a\\
&\Psi(a\tens c)=qc\tens a&\Psi(a\tens d)=d\tens a\\
&\Psi(b\tens a)=q a\tens b + (q^2-1)b\tens a& \Psi(b\tens b)=q^2 b\tens b\\
&\Psi(b\tens c)=c\tens b + (q-q^{-1})d\tens a& \Psi(b\tens d)=q d\tens b\\
&\Psi(c\tens a)=q a\tens c + (q^2-1)c\tens a& \Psi(c\tens c)=q^2 c\tens c\\
&\Psi(c\tens b)=b\tens c + (q-q^{-1})d\tens a& \Psi(c\tens d)=q d\tens c\\
&\Psi(d\tens b)=qb\tens d + (q^2-1)d\tens b&\Psi(d\tens c)=qc\tens d +
(q^2-1)d\tens c\\
&\Psi(d\tens a)=a\tens d+(q-q^{-1})(c\tens b+b\tens c)+ (q-q^{-1})^2 d\tens
a&\Psi(d\tens d)=q^2 d\tens d}

Equivalently, one can work directly with the braided tensor product algebra
$B\und\tens B$. Denoting the second copy of $A(R)$ with a prime, the algebra
$A(R)\und\tens A(R)$ is generated by $\vect,\vect'$ with their usual relations
and cross-relations from $\Psi$, namely the {\em braid statistics}
\eqn{bstat}{\vect'_1\vect_2=R_{21}\vect_2\vect'_1 R.}
The homomorphism property of $\und\Delta$ in Theorem~2.1 in this notation is
that $\vect''=\vect+\vect'$ then also obeys the FRT relations of $A(R)$.

In these terms then, let $\pmatrix{a'&b'\cr c'&d'}$ denote a second copy of
$M_q(2)$ with braid statistics
\[ a'a=q^2 aa',\quad b'b=q^2bb',\quad c'c=q^2 cc',\quad d'd=q^2 dd'\]
\[ a'b=qba',\quad a'c=qca',\quad a'd=da',\quad b'd=qdb',\quad c'd=qdc' \]
\[ b'a=qab'+(q^2-1)ba',\quad b'c=cb'+(q-q^{-1})da'\]
\[c'a=qac'+(q^2-1)ca',\quad c'b=bc'+(q-q^{-1})da'\]
\[ d'b=qbd'+(q^2-1)db',\quad d'c=qcd'+(q^2-1)dc'\]
\[ d'a=ad'+(q-q^{-1})(cb'+bc')+(q-q^{-1})^2 da'\]
Then
\[ \pmatrix{a''&b''\cr c''& d''}=\pmatrix{a&b\cr c&d}+\pmatrix{a'&b'\cr
c'&d'}\]
also obeys the relations of $M_q(2)$.

On the other hand, our construction works for the coaddition of any quantum
matrix bialgebra associated to a Hecke-type R-matrix. This includes
multiparameter and non-standard variants of the $GL_q(n)$ R-matrices and many
others.

Finally, we explain the categorical setting which is that of bicomodules. Let
us recall that in the abstract braided group theory, all structures are fully
covariant under some quantum group. This quantum group has a (dual)
quasitriangular structure which then induces the braidings, such as the
braidings above. This point of view is explained in detail in \cite{Ma:lin}.

\begin{lemma} $A(R)$ is a right $A(R)^{\rm cop}\tens A(R)$ comodule algebra.
I.e. there is a coaction $A(R)\to A(R)\tens A(R)^{\rm cop}\tens A(R)$ which is
an algebra homomorphism. Explicitly, it is given by
\[ t^i{}_j\to t^a{}_b\tens\tau^i{}_a\tens\sigma^b{}_j, \quad {\rm i.e.},\quad
\vect\to \tau\vect\sigma\]
in a compact notation. Here $A(R)^{\rm cop}$ denotes $A(R)$ with reversed
coproduct and with matrix generator $\tau$. The matrix generator of the other
coacting $A(R)$ is $\sigma$.
\end{lemma}
\proof We first note that if $A$ is a bialgebra or Hopf algebra then so is
$A^{\rm cop}$. This is the same algebra but with the reversed coproduct. Now
$A$ coacts on itself from both the left and the right via its coproduct
$\Delta:A\to A\tens A$ (the left regular and right regular coactions), and this
coproduct map is an algebra homomorphism. Now any left $A$-comodule algebra is
the same thing as a right $A^{\rm cop}$-comodule algebra. The coaction is
naturally from the left but we write the same linear map now from the right and
compensate by reversing the coproduct of $A$. This right coaction $A\to A\tens
A^{\rm cop}$ is still an algebra homomorphism. In particular then, both $A$ and
$A^{\rm cop}$ coact from the right on $A$ via the coproduct homomorphism.
Therefore $A^{\rm cop}\tens A$ also coacts from the right. This is the general
situation. Explicitly for $A=A(R)$ the coaction is as stated. In the compact
notation $\tau\vect\sigma$ we write $\tau$ on the left for the purpose of
matrix multiplication but it lives in the middle of $A(R)\tens A(R)^{\rm
cop}\tens A(R)$. One can also verify explicitly that the coaction shown extends
consistently as an algebra homomorphism to products,
\eqn{coactprod}{t^{i_1}{}_{j_1}\cdots t^{i_M}{}_{j_M}\to t^{a_1}{}_{b_1}\cdots
t^{a_M}{}_{b_M}\tens \tau^{i_1}{}_{a_1}\cdots\tau^{i_M}{}_{a_M}\tens
\sigma^{b_1}{}_{j_1}\cdots \sigma^{b_M}{}_{j_M}}
as it must from the general reasons given. It is easy to see that this is
consistent with the FRT relations $R\vect_1\vect_2=\vect_2\vect_1R$ by applying
to both sides and using these and the corresponding relations
$R\tau_1\tau_2=\tau_2\tau_1R$ and $R\sigma_1\sigma_2=\sigma_2\sigma_1R$.
\endproof

Equivalently, $A(R)$ is an $A(R)$-{\em bicomodule algebra}. For any bialgebra,
a $A$-bicomodule is by definition something on which $A$ coacts from the left
and from the right by maps $\beta_L$ and $\beta_R$ say, and these two coactions
commute in the sense
\[ (\beta_L\tens\id)\beta_R=(\id\tens\beta_R)\beta_L.\]
This is just the same thing as a right $A^{\rm cop}\tens A$-comodule since the
right $A(R)^{\rm cop}$-comodule can be viewed just as well as $\beta_L$ and the
right $A$-comodule part is $\beta_R$. This is exactly a dualisation of the
notion of bimodule for algebras. For any dual-quasitriangular bialgebra $A$,
the categories of left $A$-comodules, right $A$-comodules and hence of
$A$-bicomodules are all braided\cite{Ma:bg}.

\begin{propos} $A(R)$ with its coaddition in Theorem~2.1 is a braided-bialgebra
living in the braided category of $A(R)$-bicomodules.
\end{propos}
\proof Suppose that $A$ is a dual-quasitriangular bialgebra\cite{Ma:bg} in the
sense of a map $\CR:A\tens A\to \Bbb C$ obeying axioms dual to those of
Drinfeld\cite{Dri} for a universal R-matrix. Then $A^{\rm cop}$ is also dual
quasitriangular with $\CR^{\rm cop}(a\tens b)=\CR(b\tens a)$. Hence also
$A^{\rm cop}\tens A$ is dual-quasitriangular with the corresponding tensor
product dual quasitriangular structure,
\[ \CR\left((a\tens b)\tens (c\tens d)\right)=\CR(c\tens a)\CR(b\tens d),\qquad
\forall (a\tens b),(c\tens d)\in A^{\rm cop}\tens A.\]
So $A^{\rm cop}\tens A$ is a dual quantum group in the strict sense. In
particular, from \cite[Sec. 3]{Ma:qua} we know that $A(R)$ has dual
quasitriangular structure $\CR:A(R)\tens A(R)\to \Bbb C$ such that
$\CR(\vect_1\tens\vect_2)=R_{12}$ and extended to products as a skew
bicharacter. This is the reason that dual-quasitriangular structures were
introduced by the author (and subsequently by other authors also). Hence we
conclude that $A(R)^{\rm cop}\tens A$ has a dual-quasitriangular structure and
that therefore its category of comodule is braided. From the general scheme
explained in the present matrix context in \cite{Ma:lin}, we compute the
corresponding braiding as
\align{ &&\nqquad\Psi(t^{i_1}{}_{j_1}\cdots t^{i_M}{}_{j_M}\tens
t^{k_1}{}_{l_1}\cdots t^{k_N}{}_{l_N})\\
&&=t^{c_1}{}_{d_1}\cdots t^{c_N}{}_{d_N}\tens t^{a_1}{}_{b_1}\cdots
t^{a_M}{}_{b_M}\\
&&\qquad\qquad \CR\left((\tau^{i_1}{}_{a_1}\cdots
\tau^{i_M}{}_{a_M}\tens\sigma^{b_1}{}_{j_1}\cdots\sigma^{b_M}{}_{j_M})
\tens(\tau^{k_1}{}_{c_1}\cdots\tau^{k_M}{}_{c_M}\tens\sigma^{d_1}{}_{l_1}
\cdots\sigma^{d_N}{}_{l_N})\right)\\
&&= t^{c_1}{}_{d_1}\cdots t^{c_N}{}_{d_N}\tens t^{a_1}{}_{b_1}\cdots
t^{a_M}{}_{b_M}
Z_R({\scriptstyle \bar D\atop {{{\scriptstyle B\lform J}\atop {\scriptstyle
\bar L}}}})
Z_R({\scriptstyle \bar I\atop {{{\scriptstyle K\lform C}\atop{\scriptstyle \bar
A}}}})}
Here
\align{\CR(t^{i_1}{}_{j_1}t^{i_2}{}_{j_2}\cdots
t^{i_M}{}_{j_M}\tens t^{k_N}{}_{l_N} t^{k_{N-1}}{}_{l_{N-1}}\cdots
t^{k_1}{}_{l_1})\nqquad\nqquad\nqquad\nqquad\nqquad\nqquad\nqquad
\nqquad\nquad&&\\
&&\nquad=R^{i_1}{}_{m_{11}}{}^{k_1}{}_{n_{11}}
\ \, R^{m_{11}}{}_{m_{12}}{}^{k_2}{}_{n_{21}}\ \
\cdots \ R^{m_{1N-1}}{}_{j_{1}}{}^{k_N}{}_{n_{N1}}\\
&&\ \, R^{i_2}{}_{m_{21}}{}^{n_{11}}{}_{n_{12}}
R^{m_{21}}{}_{m_{22}}{}^{n_{21}}{}_{n_{22}}\\
&&\qquad\ \vdots\qquad\qquad\qquad\qquad\qquad\qquad\quad\vdots\cr
&&\ R^{i_M}{}_{m_{M1}}{}^{n_{1M-1}}{}_{l_1} \ \ \cdots\quad \ \, \cdots\quad
R^{m_{MN-1}}{}_{j_{M}}{}^{n_{NM-1}}{}_{l_N}
=Z_R({\scriptstyle K\atop {{{\scriptstyle I\lform J}\atop {\scriptstyle L}}}})}
is the description as a partition function\cite[Sec. 5]{Ma:qua} and
$I=(i_1,\cdots ,i_M)$ and $K=(k_1,\cdots k_N)$ while $\bar I=(i_M,\cdots,i_1)$
has the reversed order. This is the same result as $\Psi$ in the proof of
Theorem~2.1 above. \endproof

Another notation is to write $t^I{}_J=t^{i_1}{}_{j_1}\cdots t^{i_M}{}_{j_M}$
and $\CR(t^I{}_J\tens t^K{}_L)=R^I{}_J{}^K{}_L=Z_R({\scriptstyle \bar K\atop
{{{\scriptstyle I\lform J}\atop {\scriptstyle \bar L}}}})$. Then
\[ \Psi(t^I{}_J\tens t^K{}_L)=R^K{}_C{}^I{}_A\,  t^C{}_D\tens t^A{}_B\,
R^B{}_J{}^D{}_L\]
which is of the same form as on the generators in Theorem~2.1. Note that the
multi-index notation here is the one for partition functions and row-to-row
transfer matrices in statistical mechanics and not the one above that reduced
$A(R)$ to a braided covector algebra. Thus we have two ways of thinking about
$A(R)$ as living in a braided category. The two are related by the bialgebra
homomorphism
\eqn{push}{ A({\bf R})\to A(R)^{\rm cop}\tens A(R),\quad
t^{(i_0,i_1)}{}_{(j_0,j_1)}\mapsto \tau^{j_0}{}_{i_0}\tens
\sigma^{i_1}{}_{j_1}.}
By this map an $A({\bf R})$-comodule algebra in our braided-covector point of
view\cite{Ma:poi} explained directly after Theorem~2.1, pushes out to our
second $A(R)^{\rm cop}\tens A(R)$-comodule point of view.

\section{Vector fields on quantum matrices}

One thing that one can do with our new braided addition law on quantum matrices
is to make an infinitesimal addition. According to\cite{Ma:fre} this defines
braided differential operators or vector fields on the underlying braided
space. The co-ordinate functions are $\vect=(t^{i_0}{}_{i_1})=(t_I)$ so the
corresponding differentials
\[ \del^I\equiv\del^{i_1}{}_{i_0}\equiv{\del\over\del t^{i_0}{}_{i_1}}:A(R)\to
A(R)\]
defined by\cite{Ma:fre}
\[ \del^If(\vect)={}_{\veca=0}|(a_I^{-1}(f(\veca+\vect)-f(\vect)))={\rm coeff\
of}\ {a_I}\ {\rm in}\ f(\veca+\vect)\]
where $\veca=(a_I)=(a^{i_0}{}_{i_1})$ denotes the first copy of $A(R)$ in
$A(R)\und\tens A(R)$ (the second copy being denoted still by $\vect$). The two
copies have braid statistics  $\vect_1\veca_2=R_{21}\veca_2\vect_1 R$
with respect to each other from (\ref{bstat}). We assume throughout that $R$ is
of Hecke type so that the results of Section~2 apply.

These differential operators $\del^I:A(R)\to A(R)$ are constructed explicitly
in \cite{Ma:fre} in terms of braided-integers $[m,{\bf R}]$ in the form
$\del^I\vect_1\vect_2\cdots\vect_m=e^I{}_1\vect_2\cdots\vect_m[m:{\bf R}]$
where ${\bf R}$ from Section~2 is the relevant braiding matrix for linear
addition. Because they correspond to linear addition, they are the analogues of
the usual cartesian partial derivatives $\del\over\del t^i{}_j$.

On the other hand, on a group or matrix space, there are also the
left-invariant vector fields $\tilde\del^I$ say given by right-multiplication
in the underlying ring of matrices. These are constructed in terms of the usual
coproduct $\Delta$ corresponding to the multiplication. They are related to the
cartesian $\del^I$ via the codistributivity property in Definition~2.2. Our
goal in this section is to compute this relationship explicitly. We will also
see that they form some kind of braided-Lie algebra.

We begin by computing the algebra generated by $\del^I$ and its action on
$A(R)$.

\begin{propos} The braided-differential operators $\del^I=\del^{i_1}{}_{i_0}$
obey the relations $\del_1\del_2R=R\del_2\del_1$ of $A(R_{21})$.
\end{propos}
\proof In the general scheme of \cite{Ma:fre} the braided derivatives obey the
braided vector algebra $V({\bf R'})$ with relations and braiding
\eqn{vect}{\Psi(\del^I\tens \del^J)={\bf R}^I{}_A{}^J{}_B\del^B\tens
\del^A,\quad \del^I\del^J={\bf R'}^I{}_A{}^J{}_B\del^B\del^A}
for $\bf R$ and $\bf R'$ as for the covector algebra
(\ref{psimulti})--(\ref{A(R)multi}).  This too forms a braided group with
linear addition law, and is in some sense the dual of the covector braided
group. Putting in the form of ${\bf R'}$ we have
\[ \del^{i_1}{}_{i_0}\del^{j_1}{}_{j_0}=R^{-1}{}^{a_0}{}_{i_0}{}^{b_0}{}_{j_0}
R^{i_1}{}_{a_1}{}^{j_1}{}_{b_1}\del^{b_1}{}_{b_0}\del^{a_1}{}_{a_0}\]
which gives the relations shown. \endproof

Both the vectors and the covectors live in a braided category hence there are
braidings among them too. These are given in \cite{Ma:lie} in the general case,
and we will need them explicitly
\eqn{bra-td}{\Psi(t_I\tens \del^J)=\widetilde{\bf R}^A{}_I{}^J{}_B\del^B\tens
t_A,\quad \widetilde{\bf
R}^A{}_I{}^J{}_B=\tilde{R}^{b_0}{}_{j_0}{}^{i_0}{}_{a_0}
\tilde{R}^{a_1}{}_{i_1}{}^{j_1}{}_{b_1}}
\eqn{bra-dt}{ \Psi(\del^I\tens t_J)={\bf R}^{-1}{}^I{}_A{}^B{}_Jt_B\tens
\del^A,\quad {\bf
R}^{-1}{}^I{}_A{}^B{}_J=R^{-1}{}^{j_0}{}_{b_0}{}^{a_0}{}_{i_0}
R^{-1}{}^{i_1}{}_{a_1}{}^{b_1}{}_{j_1}}
where $\tilde R$ denotes the second inverse $((R^{t_2})^{-1})^{t_2}$ and $t_2$
is transposition in the second matrix factor. In particular, the $\del^I$ obey
a braided-Leibniz rule for computing $\del^I(f(\vect)g(\vect))$ whereby
$\del^I$ is taken past $f(\vect)$ with a $\Psi^{-1}$\cite{Ma:fre}. This is the
generalisation of the philosophy of super-differentiation.

Next we compute the braided Heisenberg-Weyl algebra, which is the algebra of
$R$-quantum mechanics generated by the vectors $\del^I$ acting on the
co-ordinate functions $t_J$. The general scheme in the braided setting is in
\cite{Ma:fre} and gives the usual relations
\[ \del^I t_J-t_A {\bf R}^A{}_J{}^I{}_B\del^B=\delta^I{}_J\]
which in our case becomes
\eqn{weyl}{\del^{i_1}{}_{i_0} t^{j_0}{}_{j_1}-R^{a}{}_{i_0}{}^{j_0}{}_{b}
t^{b}{}_{c}
R^{c}{}_{j_1}{}^{i_1}{}_{d}\del^{d}{}_{a}
=\delta^{i_1}{}_{j_1}\delta^{j_0}{}_{i_0}.}
The structure here is that of a braided-semidirect product\cite{Ma:fre} which
we see is $A(R)\cocross A(R_{21})$. It is built on $A(R)\tens A(R_{21})$ with
the cross relations (\ref{weyl}).

In this Heisenberg-Weyl algebra are the vector fields on $A(R)$ as the subspace
of the form $f_I(\vect)\del^I$. One can usually define on this set some form a
braided Lie bracket corresponding to the Lie derivative of vector fields, i.e.
some form of a diffeomorphism Lie algebra. This includes the case of the
q-Virasoro Lie algebra but, as for this, the coalgebra and the general picture
are both unknown. On the other hand, certain vector fields such as the
left-invariant vector fields on a group or matrix space should close under the
Lie bracket into some form of sub-braided Lie algebra similar to the
corresponding quantum enveloping algebra in the standard cases.

We recall first the classical formula for the left-invariant vector fields on a
matrix group $G$. Let $\xi$ be in the Lie algebra of $G$. The corresponding
left-invariant vector field is defined on $f\in C(G)$ by
\[ (\tilde\xi\la f)(g)={d\over d\eps}|_0 f(g(1+\eps\xi))={d\over
d\eps}|_0\left( f(g)+\eps (g\xi)^i{}_j{\del\over\del g^i{}_j}
f(g)\right)=g^i{}_a\xi^a{}_j{\del\over\del g^i{}_j} f(g).\]
If we define $L_h^*$ by left translation in the group as $L^*_h(f)(g)=f(hg)$
for $h,g\in G$ then
\[ (L^*_h(\tilde\xi f))(g)=(\tilde\xi f)(hg)={d\over d\eps}|_0
f(hg(1+\eps\xi))={d\over d\eps}|_0 (L_h^*f)(g(1+\eps\xi))=(\tilde\xi (L_h^*
f))(g)\]
which expresses left-invariance of the $\tilde\xi$.

Abstractly in terms of the co-ordinate functions $\vect$ we analogously define
the {\em left-invariant vector fields}
\eqn{left-vec}{ \tilde\xi f(\vect)=t^i{}_a\xi^a{}_j{\del\over\del
t^i{}_j}f(\vect),\quad {\rm i.e.} \quad
\tilde\xi=\xi^{i_0}{}_{i_1}\tilde\del^{i_1}{}_{i_0},\qquad
\tilde\del^I\equiv\tilde\del^{i_1}{}_{i_0}=t^a{}_{i_0}\del^{i_1}{}_a}

\begin{propos} The vector fields $\tilde\del^I:A(R)\to A(R)$  defined in
(\ref{left-vec}) are left-invariant in the sense
\[ (\id\tens\tilde\del^I)\Delta=\Delta\tilde\del^I.\]
\end{propos}
\proof This follows from the codistributivity properties proven in
Proposition~2.3. We first introduce the functionals
\[ \eps^I\equiv\eps^{i_1}{}_{i_0},\qquad
\eps^I(t_J)=\delta^I{}_J=\delta^{i_1}{}_{j_1}\delta^{j_0}{}_{i_0}\]
with $\eps^I=0$ on all other powers of $\vect$. With respect to the usual
multiplicative coproduct $\Delta$ it obeys
\[ (\id\tens\eps^I)\Delta=t^a{}_{i_0}\eps^{i_1}{}_a \]
since both sides are zero except when acting on anything of the form
$t^a{}_{i_1}$. For each $a$, the coproduct $\Delta t^a{}_{i_1}$ contains the
term $t^a{}_{i_0}\tens t^{i_0}{}_{i_1}$ which is the only contribution to the
left hand side. Then both sides give the same result. In terms of this linear
functional $\eps^I$ we have the definition of $\del^I$ as
\[ \del^I=(\eps^I\tens\id)\und\Delta.\]
Now we can compute
\align{(\id\tens\del^I)\Delta&=&
(\id\tens\eps^I\tens\id)(\id\tens\und\Delta)\Delta\\
&=&(\id\tens\eps^I\tens\id)(\cdot\tens\id\tens\id)\tau_{23}
(\Delta\tens\Delta)\und\Delta\\
&=&(\cdot\tens\id)(\id\tens\id\tens\eps^I\tens\id)\tau_{23}
(\Delta\tens\Delta)\und\Delta\\
&=&(\cdot\tens\id)((\id\tens\eps^I)\Delta\tens\Delta)\und\Delta\\
&=&(\cdot\tens\id)(\id\tens\Delta)(
t^a{}_{i_0}\eps^{i_1}{}_a\tens\id)\und\Delta\\
&=&(\cdot\tens\id)(\id\tens\Delta) t^a{}_{i_0}\tens\del^{i_1}{}_a\\
&=& (t^a{}_{i_0}\tens 1)\Delta\circ\del^{i_1}{}_a}
where the first equality is our new form for $\del^I$, the second equality is
codistributivity, the third and fourth are rearrangements, the fifth is our
property for $\eps^I$ above in relation to $\Delta$. We then rearrange to write
in terms of $\del$ again.

This result for $\del^I$ then implies
\[(\id\tens t^b{}_{i_0}\del^{i_1}{}_b)\Delta=(t^a{}_b\tens
t^b{}_{i_0})\Delta\del^{i_1}{}_a =\Delta\circ t^a{}_{i_0}\del^{i_1}{}_a\]
which is the left-invariance result for $\tilde\del$. \endproof

On any bialgebra or Hopf algebra $A$ one has the left regular representation of
$A^*$ defined by $h\la a=\sum a\o <a\t, h>$ where $\Delta a=\sum a\o\tens a\t$
and $<\ ,\ >$ is the pairing. These operators $A\to A$ are also left-invariant
in the sense $\Delta(h\la a)=(\id\tens h\la)\Delta a$ as above. Conversely,
every left-invariant  operator is of this form for some $h\in A^*$. Given a
left-invariant operator $D$, the required linear functional is $h=\eps\circ D$
where $\eps$ is the counit.

\begin{propos} Our left-invariant differential operators $\tilde\del^I$
correspond to linear functionals $L^I\in A(R)^*$ given by $L^i{}_j(1)=0$ and
\align{ &&\nquad L^i{}_j(t^{i_1}{}_{j_1}\cdots
t^{i_m}{}_{j_m})=\delta^i{}_{j_1}\delta^{i_1}{}_j\delta^{i_{2}}{}_{j_{2}}
\cdots\delta^{i_{m}}{}_{j_m} + R^{i_2}{}_j{}^{i_1}{}_{c}
R^{c}{}_{j_1}{}^i{}_{j_2}\delta^{i_{3}}{}_{j_{3}}\cdots\delta^{i_{m}}{}_{j_m}\\
&&+\sum_{r=2}^m R^{i_r}{}_{a_{r-2}}{}^{i_{r-1}}{}_{c_{r-1}}\cdots
R^{a_2}{}_{a_1}{}^{i_2}{}_{c_2} R^{a_1}{}_j{}^{i_1}{}_{c_1}
R^{c_1}{}_{j_1}{}^i{}_{b_1} R^{c_2}{}_{j_2}{}^{b_1}{}_{b_2} \cdots
R^{c_{r-1}}{}_{j_{r-1}}{}^{b_{r-2}}{}_{j_r}\,
\delta^{i_{r+1}}{}_{j_{r+1}}\cdots\delta^{i_{m}}{}_{j_m}.}

\end{propos}
\proof The action of $\del^I$ on monomials is given in \cite{Ma:fre} in terms
of $\bf R$. We write this in terms of $R$ as in (\ref{psimulti}), convert to
$\tilde\del^I$ by another factor $\vect$ and then evaluate against the counit.
Thus we compute
\[L^i{}_j(t_{I_1}\cdots t_{I_m})=
\eps\left(t^a{}_j\delta^{(a,i)}{}_{J_1}t_{J_2}\cdots t_{J_m}[m,{\bf
R}]^{J_1\cdots J_m}_{I_1\cdots I_m}\right)\]
where $J=(j_0,j_1)$ etc as before. The braided integers are defined in
\cite{Ma:fre} by $[m;{\bf R}]=1+P{\bf R}_{12}+\cdots + P{\bf R}_{12}P{\bf
R}_{23}\cdots P{\bf R}_{m-1m}$. Putting the form of $\bf R$ into this and
evaluating with $\eps(t_I)=\delta_I=\delta^{i_0}{}_{i_1}$ gives the result
shown after conversion to another notation for the numbering of the indices.
\endproof

This is as a pair of single-row transfer matrices. It seems likely that at
least for the standard $R$-matrices these linear functionals restrict to the
group function algebras and can be written in terms of the FRT generators
$l^\pm$ for the quantum enveloping algebra. We have not found any general
formula of this form, however.

Now we study the algebra generated by these $\tilde\del^I$ inside the braided
Heisenberg-Weyl algebra. By definition any closed commutation relations inside
here are to be thought of as restrictions of some braided-Lie derivative. In
general we expect the $\tilde\del^I$ to close and form a braided-Lie algebra
version of $A^*$ via the correspondence in the last proposition.

Before describing this, we recall the braided-matrix algebras $B(R)$ introduced
in \cite{Ma:exa}. This works for any bi-invertible $R$-matrix (not necessarily
Hecke) and has a matrix of generators $\vecu=(u^{i_0}{}_{i_i})=(u_I)$ with
braiding and relations as follows. To be compatible with the left-handed
conventions above it turns out we need $B(R_{21})$ rather than $B(R)$. Then
\[ \Psi(u_I\tens u_J)=u_B\tens u_A \Psi^B{}_J{}^A{}_I,\quad
\Psi^I{}_K{}^J{}_L=R^c{}_{i_0}{}^{l_0}{}_d R^{-1}{}^{i_1}{}_a{}^d{}_{j_0}
R^a{}_{k_1}{}^{j_1}{}_b\tilde{R}{}^{k_0}{}_c{}^b{}_{l_1}\]
\[u_Iu_J=u_Bu_A\Psi'{}^B{}_J{}^A{}_I,\quad
\Psi'{}^I{}_K{}^J{}_L=R^{-1}{}^{l_0}{}_{d}{}^{c}{}_{i_0}
R^{d}{}_{j_0}{}^{i_1}{}_{a}
R^a{}_{k_1}{}^{j_1}{}_b\tilde{R}{}^{k_0}{}_c{}^b{}_{l_1}.\]
These matrices are taken from \cite{Ma:exa} with some minor rearrangement of
conventions for our present purposes.  The relations and braiding can also be
written in the compact form $R\vecu_1R_{21}\vecu_2=\vecu_2R\vecu_1R_{21}$ and
$R\vecu'_1R^{-1}\vecu_2=\vecu_2R\vecu'_1R^{-1}$. The first of these is also
known in other contexts, while the second arises only in the theory of braided
groups. Our point of view in \cite{Ma:exa} was  as defining the
braided-commutative algebra of functions on a braided version of $M_n$. Thus
$\Psi$ obeys the QYBE and $\Psi'$ is a variant of it so that the relations are
like the super-commutativity for the functions on a super-space.

\begin{propos} The braiding between $\tilde\del^I$ and their relations inside
the braided Heisenberg-Weyl algebra are
\[\Psi(\tilde\del^I\tens\tilde\del^J)=\Psi^I{}_A{}^J{}_B\tilde\del^B\tens
\tilde\del^A,\quad\tilde\del^I\tilde\del^J-\Psi'{}^I{}_A{}^J{}_B\tilde\del^B
\tilde\del^A=\tilde\del^{j_1}{}_{i_0}\delta^{i_1}{}_{j_0}-\Psi'{}^I{}_A{}^J{}_B
\tilde\del^{a_1}{}_{b_0}\delta^{b_1}{}_{a_0}\]
\end{propos}
\proof The braidings are computed from (\ref{psimulti}), (\ref{vect}) and
(\ref{bra-td})-(\ref{bra-dt})  for the braidings between $\del,\vect$. To braid
$t^a{}_{i_0}\tens\del^{i_1}{}_a$ past  $t^b{}_{j_0}\tens\del^{j_1}{}_b$ first
braid $\del^{i_1}{}_a$ past $t^{b}{}_{j_0}$, then the $\vect$ resulting from
this past $\del^{j_1}{}_b$. Then braid $t^{a}{}_{i_0}$ to the right in the same
way. The result is the braiding shown.

The computation for the relations is similar, using this time the relations of
$A(R)$ for $\vect$, of $A(R_{21})$ for $\del$ and the Heisenberg-Weyl relations
(\ref{weyl}) between $\vect$ and $\del$. Between the $\vect$ and $\tilde\del$
the relations come out as
\[ \tilde\del^{i_1}{}_{i_0}t^{j_0}{}_{j_1}=\delta^{i_1}{}_{j_1}t^{j_0}{}_{i_0}+
t^{j_0}{}_a\tilde\del^b{}_c R^c{}_{i_0}{}^a{}_d R^d{}_{j_1}{}^{i_1}{}_b\]
\[ t^{i_0}{}_{i_1}\tilde\del^{j_1}{}_{j_0}=\tilde{R}^a{}_b{}^{j_1}{}_c
\tilde\del^c{}_d t^{i_0}{}_{a}R^{-1}{}^d{}_{j_0}{}^b{}_{i_1}-t^{i_0}{}_a
\tilde{R}^c{}_b{}^{j_1}{}_c R^{-1}{}^a{}_{j_0}{}^b{}_{i_1}\]
The second of these is found by applying $\tilde R$ and $R^{-1}$ to both sides
of the first. Using these relations one finds easily the result stated.
\endproof

The classical limit of these constructions is with $R=\id$. Then the relations
between the $\tilde\del^I$ become
\[[\tilde\del^{i_1}{}_{i_0},\tilde\del^{j_1}{}_{j_0}]=\tilde\del^{j_1}{}_{i_0}
\delta^{i_1}{}_{j_0}-\delta^{j_1}{}_{i_0}\tilde\del^{i_1}{}_{j_0}\]
which is how the Lie algebra $gl_n$ is realised as left-invariant vector fields
on the functions on $M_n$. Our constructions are exactly a braided version of
this. On the other hand, we have not found a suitable braided coalgebra
structure such that the relations in the proposition form a braided Lie algebra
in the strict sense of \cite{Ma:lie}. Nevertheless, the two approaches are
closely related and perhaps equivalent.

For our canonical example, we take the $GL_2$ R-matrix (\ref{Rgl2}) as in
Section~2 and write
\[[\tilde\del^I,\tilde\del^J]_{\Psi'}\equiv
\tilde\del^I\tilde\del^J-\Psi'{}^I{}_A{}^J{}_B\tilde\del^B\tilde\del^A,\qquad
(\tilde\del^i{}_j)\equiv\pmatrix{\alpha &\beta\cr \gamma & \delta}.\]
Then the $gl_2$ braided-Lie algebra-like structure from Proposition~3.4 comes
out explicitly as
\[ [\alpha,\beta]_{\Psi'}=-\beta=-[\beta,\alpha]_{\Psi'},\quad
[\alpha,\gamma]_{\Psi'}=q^{-2}\gamma=-[\gamma,\alpha]_{\Psi'},\quad
[\beta,\gamma]_{\Psi'}=\delta-q^{-2}\alpha=-[\gamma,\beta]_{\Psi'}\]
\[[\alpha,\delta]_{\Psi'}=(q^{-2}-1)(\delta-q^{-2}\alpha)=-[\delta,
\delta]_{\Psi'},\quad[\delta,\alpha]_{\Psi'}=0\]
\[ [\beta,\delta]_{\Psi'}=-q^{-4}\beta=- q^{-4}[\delta,\beta]_{\Psi'},\quad
[\gamma,\delta]_{\Psi'}=\gamma (1+q^{-2}-q^{-4}),\quad
[\delta,\gamma]_{\Psi'}=-q^{-2}\gamma\]
with the remaining three zero. These can be compared with the braided-Lie
algebra $\und gl_2$ in \cite{Ma:lie}.

Meanwhile, the relations themselves from Proposition~3.4 compute as
\eqn{bgl2a}{ \alpha\beta-q^{-2}\beta\alpha=-q^{-2}\beta,\qquad
\alpha\gamma-q^2\gamma\alpha=\gamma}
\eqn{bgl2b}{\beta\gamma-\gamma\beta=(1+\alpha(q^2-1))(\delta-q^{-2}\alpha),
\qquad\alpha+\delta\quad {\rm central}.}
The braiding is computed from $\Psi$ and is similar to that for the $\und gl_2$
generators. We do not need it directly in the algebra. The algebra-relations
themselves are an extension by linear terms of a variant of the quadratic
braided-matrices algebra  $BM_q(2)=U(\und gl_2)$. To see this note that one may
rescale the $\tilde\del$ so that a new parameter $\hbar$ appears in front of
all the linear terms above. Then the limit $\hbar\to 0$ recovers a variant of
these algebras, namely an algebra isomorphic to $BM_{q^{-1}}(2)$ after a change
and
rescaling of the generators. The braided-matrices algebra was studied
extensively in \cite{Ma:exa} where we explained the sense in which its
relations were those of `braided-commutativity'. The equations above now add a
linear right hand side to the corresponding braided-commutatator.

\section{Coaddition on rectangular quantum matrices}

In this section we extend the results of the last two sections to coaddition
and differentiation for rectangular quantum matrices. For $m\times 1$ or
$1\times n$ we recover the addition law and differential calculus for quantum
planes in \cite{Ma:poi}\cite{Ma:fre}. For square $n\times n$ matrices we
recover of course the results above. Most importantly however, all of these are
now unified into a single linear-differential calculus. Thus, as an application
we compute the vector fields (orbital angular momentum operators) for the
coaction of a quantum matrix group on a quantum space. From this point of view,
the $1\times n$ row vector is the $q$-deformed position co-ordinate vector
$\vecx=(x_i)$.

The general $m\times n$ quantum matrices also have a physical interpretation as
follows. We can think of each row as a particle position co-ordinate and the
entire matrix as a lattice model of a trajectory $\{x_i(t)\}$ say. Then the
ability to add rectangular quantum matrices is crucial and in turn defines the
notion $\delta\over\delta x_i(t)$ of functional differentiation. Likewise, one
can view a wave-function on space as a row vector $\psi(x)$ or a general
vector-valued field as a rectangular quantum matrix $\{ \psi^\alpha(x)\}$. If
one is serious about q-deforming physical constructions, one needs to be able
to add such fields pointwise, as well as to functionally differentiate with
respect to them. Hence the results in this section are the first and most basic
steps towards a systematic q-deformed or braided classical field theory and a
theory of q-deformed or braided path-integration. One may also expect plenty of
other applications of our basic notions of addition and differentiation.

We recall first the definition of rectangular quantum matrices as introduced
previously in \cite{MaMar:glu} in the theory of block decomposition of quantum
matrices into quantum blocks. The idea behind the definition is to use two
independent solutions $R_1, R_2$ say of the QYBE, one for the rows and one for
the columns. They can be of different dimensions. The associated quantum
matrices may be (co)-multiplied whenever the columns of one matches the rows of
the other, not only in dimension (as usual for matrices) but in flavour of QYBE
solution also. The $R_i$ need not in fact obey the QYBE but that is the case of
most immediate interest in the present paper.

Given $R_1\in M_m\tens M_m$ and $R_2\in M_n\tens M_n$, the associated
quantum matrix algebra $A(R_1:R_2)$ has a matrix of generators
$\vecx=(x^\mu{}_i)$ where greek indices run $\mu=1,\cdots,m$ and latin
ones $i=1,\cdots, n$, and relations
\eqn{rectmat}{R_1\vecx_1\vecx_2=\vecx_2\vecx_1R_2,\quad {\rm
i.e.\ }R_1^\mu{}_\alpha{}^\nu{}_\beta x^\alpha{}_i x^\beta{}_j=x^\nu{}_b
x^\mu{}_a
R_2^a{}_i{}^b{}_j.}
The multiplication between compatible rectangular matrices is expressed
now as a family of algebra homomorphisms\cite{MaMar:glu}
\eqn{rectcoprod}{
\Delta_{R_1,R_2,R_3}:A(R_1:R_3)\to A(R_1:R_2)\tens A(R_2:R_3)}
for any three matrices $R_i$. The map is given by the matrix coproduct
of the relevant generators, $\Delta x^\mu{}_s=x^\mu{}_a\tens
x^a{}_s$ where $s$ runs in the range appropriate for $R_3$.
Moreover, the family of maps {\em taken together} are coassociative in
the sense\cite{MaMar:glu}
\eqn{rect-coassoc}{ ({\rm id}\tens
\Delta_{R_2,R_3,R_4})\circ\Delta_{R_1,R_2,R_4}=(\Delta_{R_1,R_2,R_3}\tens{\rm
id})\circ\Delta_{R_1,R_3,R_4}}
as a map $A(R_1:R_4)\to A(R_1:R_2)\tens A(R_2:R_3)\tens A(R_3:R_4)$.
This includes all possible coassociativity conditions arising from
associativity of rectangular matrix multiplication. In summary,
rectangular quantum matrices are not individually bialgebras but they
all fit together into a weaker `co-groupoid' structure on the entire
family.

These maps include both coproducts for rectangular matrices, when
(\ref{rect-coassoc}) reduces to
usual coassociativity, and matrix coactions
\[ \beta_L: A(R_1:R_2)\to A(R_1)\tens A(R_1:R_2),\quad \beta_R: A(R_1:R_2)\to
A(R_1:R_2)\tens A(R_2)\]
corresponding to matrix multiplication of a rectangular matrix by a square
quantum matrix from the left or the right. Then (\ref{rect-coassoc}) reduces to
the comodule property for these maps.

Since we can add elements of quantum planes via a
braided-coaddition\cite{Ma:poi}, it is natural to do this too for general
rectangular or square quantum matrices. We have covered the most important
square case in Section~2 and now we summarise the extension to the rectangular
case. It was announced in \cite{MaMar:glu}. From now on, we assume that all
matrices $R_i$ are solutions of the QYBE and of Hecke type.

\begin{propos} Let $R_i$ be Hecke solutions of the QYBE. Then the rectangular
quantum matrices $A(R_1:R_2)$ form a braided-Hopf algebra
\[ \Psi(\vecx_1\tens\vecx_2)=(R_1)_{21}\vecx_2
\tens\vecx_1R_2,\quad \und\Delta_{R_1,R_2}\vecx=\vecx\tens1+1\tens\vecx,\quad
\und\eps{\vecx}=0.\]
Moreover, the coaddition is codistributive with respect to
the product of quantum matrices in the sense
\[ ({\rm
id}\tens\cdot)\circ(\id\tens \tau\tens\id)\circ(\Delta_{R_1,R_2,R_3}\tens
\Delta_{R_1,R_2,R_3})\circ\und\Delta_{R_1,R_3}=(\und\Delta_{R_1,R_2}\tens{\rm
id})\circ\Delta_{R_1,R_2,R_3}\]
\[ (\cdot\tens {\rm
id})\circ(\id\tens \tau\tens\id)\circ(\Delta_{R_1,R_2,R_3}\tens
\Delta_{R_1,R_2,R_3})\circ\und\Delta_{R_1,R_3}=({\rm id}\tens
\und\Delta_{R_1,R_2})\circ\Delta_{R_1,R_2,R_3}\]
where $\tau$ denotes usual transposition.
\end{propos}
\proof The detailed proof follows the same strategy as in the proofs of
Theorem~2.1 and Proposition~2.3 in Section~2. The refinement is to allow the
indices in the matrix products to run over their relevant ranges and to use the
corresponding $R$-matrices. \endproof

We can also regard $A(R_1:R_2)$ as a single quantum covector space as in
(\ref{psimulti}) but now
\eqn{rectmulti}{{\bf R}^A{}_I{}^B{}_J=R_1^{j_0}{}_{b_0}{}^{i_0}{}_{a_0}
R_2^{a_1}{}_{i_1}{}^{b_1}{}_{j_1},\quad
{\bf R}'{}^A{}_I{}^B{}_J=R_1^{-1}{}^{i_0}{}_{a_0}{}^{j_0}{}_{b_0}
R_2^{a_1}{}_{i_1}{}^{b_1}{}_{j_1}.}
The conventions are with $\vecx_I=x^{i_0}{}_{i_1}$ so all $0$-subscripted
indices run over the range for $R_1$ and all $1$-subscripted indices over the
range for $R_2$. Finally, it is easy to see that $A(R_1:R_2)$ lives as a
braided-Hopf algebra in the braided category of $A(R_1)-A(R_2)$--bicomodules
via left, right coactions $\beta_L,\beta_R$.
This category is also that of right comodules of $A(R_1)^{\rm cop}\tens A(R_2)$
which is a dual-quasitriangular bialgebra. The two points of view are connected
along the lines of (\ref{push}).

This summarises the rectangular version of Section~2, which results are
recovered now as special cases. But we also recover just as well the coaddition
on the quantum planes and compatibility (`colinearity') of the coaction of the
corresponding quantum matrix groups on them as a new special case. This will be
our main application in the present section. Note that in our unified approach
the one-dimensional $R$-matrices $(q)$ and $(-q^{-1})$ are perfectly good
solutions of the QYBE of Hecke type. Then $A(R:q)$ and $A(R:-q^{-1})$ are two
quantum column vector algebra, equipped with natural left-coactions of $A(R)$.
Likewise $A(q:R)$ and $A(-q^{-1}:R)$ are two quantum row vector algebras
equipped with natural right-coactions of $A(R)$. This includes the known facts
about quantum planes associated to Hecke $R$-matrices into a single framework.

The results of Section~3 likewise generalise just as easily. Regarding
$\{x_I\}$ as quantum $mn$-dimensional covector space with braiding and
relations from (\ref{rectmulti}) we apply \cite{Ma:fre} and have at once
rectangular $n\times m$-matrices $\del^i_\mu={\del\over\del x^\mu{}_i}$ or in
the multi-index notation
\[ \del^I=\del^{i_1}{}_{i_0}:A(R_1:R_2)\to A(R_1:R_2)\]
defined in the same way as an infinitesimal left translation. It sends
$f(\vecx)$ to the  coefficient of $a^{i_0}{}_{i_1}$ in $f(\veca+\vecx)$ where
$\veca$ is another rectangular quantum matrix with statistics
$\vecx_1\veca_2=(R_1)_{21}\vecx_2\veca_1 R_2$. This time a computation along
the lines of Proposition~3.1 gives that $\del^I$ obey the relations
$\del_1\del_2R_1=R_2\del_2\del_1$ of $A((R_2)_{21}:(R_1)_{21})=A(R_2:R_1)^{\rm
op}$. The general rule is to use the same formulae as in Section~3 but with
$R_1$ or $R_2$ chosen according to the flavour of the indices. The same pattern
applies to the braidings between $\del^I$ and $x_J$ in
(\ref{bra-dt})-(\ref{bra-td}). Finally, the braided Heisenberg-Weyl algebra
(\ref{weyl}) generalises in the same way as
\eqn{rectweyl}{\del^{i}{}_{\mu}x^{\nu}{}_{j}-R_1^{\alpha}{}_{\mu}
{}^{\nu}{}_{\beta}x^{\beta}{}_{a}R_2^{a}{}_{j}{}^{i}{}_{b}\del^{b}{}_{\alpha}
=\delta^{i}{}_{j}\delta^{\nu}{}_{\mu}.}
The elements $f_I(\vecx)\del^I$ are the quantum vector fields on $A(R_1:R_2)$.

Consider now such a quantum matrix algebra $A(R_1:R)$ as a right
$A(R)$-comodule algebra by $\beta_R$ corresponding to  multiplication from the
right of an $m\times n$ quantum matrix $\vecx$ by an $n\times n$ quantum matrix
$\vect$. This coaction induces quantum-vector fields on $A(R_1:R)$ in just the
same fashion as the left-invariant vector fields in Section~3. The basis of
these vector fields is provided now by
\[ \tilde\del^{i}{}_j=x^\alpha{}_j\del^i{}_\alpha:A(R_1:R)\to A(R_1:R).\]
These operators commute with the left $A(R_1)$-comodule algebra structure
$\beta_L$ corresponding to multiplication from the left by a $m\times m$
quantum matrix. Thus they are left-invariant in the sense
\[ (\id\tens\tilde\del^i{}_j)\beta_L=\beta_L\tilde\del^i{}_j.\]
Thus the construction of Proposition~3.2 generalises to the rectangular case.
The proofs are strictly analogous.

Finally, the braiding and relations among the quantum vector fields
$\tilde\del^i{}_j$ can be computed just as in Section~3. They come out exactly
as in Proposition~3.4 (independently of $R_1$). The reason is that while the
intermediate computations for $\tilde\del^i{}_j x^\mu{}_k$ etc involve both $R$
and $R_1$, the resulting formulae for the braiding and relations between
$\tilde\del$ involve only the latin indices associated to $R$. Thus we see that
exactly the same braided-Lie algebra structure which acted in Proposition~3.4
as left-invariant vector fields on $A(R)$ now acts as $\beta_L$-invariant
vector fields on $A(R_1:R)$.

\begin{corol} Let $R_1=(q)$ and $R$ a Hecke solution of the QYBE. Then $A(q:R)$
is the covector algebra with generators $\vecx=(x_i)$ and relations
$q\vecx_1\vecx_2=\vecx_2\vecx_1R$. The partial derivatives $\del=(\del^i)$ obey
the relations $\del_1\del_2 q=R\del_2\del_1$ of the vector algebra
$A(R_{21},q)$. Inside the Heisenberg-Weyl algebra $\del^ix_j-x_a
qR^a{}_j{}^i{}_b\del^b=\delta^i{}_j$ one has the vector fields for the coaction
$\vecx\to \vecx\vect$ of $A(R)$. They are given by
\[ \tilde\del^i{}_j=x_j\del^i\]
and obey the braided-Lie algebra relations and braiding in Proposition~3.4.
\end{corol}
\proof The first part reduces, as it must, to the differential calculus for
quantum planes in \cite{Ma:fre}. This part is not limited to the Hecke case.
The second part concerning $\tilde\del^i{}_j$ is our new result and follows at
once from the theory above. One can also verify it directly. Note that if we
introduce a $\hbar$ on the right hand side of the Heisenberg-Weyl algebra, it
appears also on the right hand side of the Lie-algebra-like relations in
Proposition~3.4. Moreover, sending $\hbar\to 0$ then recovers a variant of
Proposition~3.5 of \cite{Ma:lin}, of which this corollary is therefore a
generalisation. \endproof

For our canonical example with $R$-matrix (\ref{Rgl2}) we have for
$\vecx=(x,y)$ the usual quantum plane $yx=qxy$. Its partial derivatives
$\del=({\del\over\del x},{\del\over\del y})$ were recovered in our general
scheme as the operators on polynomials in $x,y$ given in Example~2.4 of
\cite{Ma:fre}. Using the formulae there, one may easily verify that
\[ \alpha=\tilde\del^1{}_1=x{\del\over\del
x},\quad\beta=\tilde\del^1{}_2=y{\del\over\del x},\quad
\gamma=\tilde\del^2{}_1=x{\del\over\del y},\quad
\delta=\tilde\del^2{}_2=y{\del\over\del y}\]
obey the relations (\ref{bgl2a})--(\ref{bgl2b}) as they must according to the
above corollary.

These quantum vector fields are like orbital angular momentum because they are
induced by the coaction of $A(R)$ and are indeed the action of some braided
version of $gl_n$ as explained at the end of Section~3. Such  orbital angular
momentum  realisations of quantum enveloping algebras or their braided-Lie
algebras can surely be constructed by hand in low dimensions. However, we have
derived them here in an entirely systematic way and one that works for all
dimensions and general (Hecke) R-matrices.  We note that some physically
relevant cases such as rotations in $1+3$ dimensions can be expected to follow
along broadly similar lines but require us to leave the Hecke setting. This is
a topic for further work.

Finally, we recall a different kind of possible application. This is to view
infinite-dimensional row or column matrices as fields or trajectories, as
mentioned above. This time we could consider for example that
$x^\mu{}_i=x_i(\mu)$ where $\mu$ is a discretised time variable. Matrix
multiplication from the left would now be convolution of a time-dependent
function against a matrix kernel. Moreover, our derivatives
$\del^i{}_\mu={\delta\over\delta x_i(\mu)}$ become functional derivatives. The
point is that the superposition principle ensures that many constructions in
field theory are nothing more than infinite-dimensional linear algebra. When
discretised one has in the q-deformed setting many copies (one at each lattice
site) of some non-commutative algebra. One also has in general the need for
non-commutation relations between the sites. In this case one needs a
systematic formulation of such algebras and one that expresses linearity and
other familiar properties. The general theory of rectangular quantum matrices
achieves this at least in the $GL_n$ or Hecke setting. For example, a classical
but q-deformed trajectory in one dimension could be formulated as living in
$A(q:R_\infty)$ where $R_\infty$ is the $GL_\infty$ $R$-matrix. Discretised
wave-functions too can in principle be formulated this way. Some detailed
applications of this point of view will be developed elsewhere, perhaps in
relation to \cite{Kem:sym}. For the present we limit ourselves to a remark
about quantum loop groups arising from this general point of view. This is the
topic of the next section.

\section{Remark on lattice model of Kac-Moody Lie algebras}

We conclude by explaining how the notion of rectangular quantum matrix studied
in the last section leads naturally to the algebra of the discretised
wave-functions on in the lattice approximation to Kac-Moody algebras in
\cite{AFS:hid}. We learn that we can add such fields, as well as differentiate
with respect to them. There remarks then can be viewed as a small step towards
a braided-geometrical picture of a Kac-Moody algebra as a q-deformed or braided
loop-group algebra.

Thus, let $G$ be a group and $LG$ the group of maps or trajectories $S^1\to G$
with pointwise product. Now $C(LG)$ means functions on the space of such maps,
i.e. in a discrete approximation it means $C(G\times G\cdots\times G)$. The
q-deformed version of this is therefore $A(R)\und\tens
A(R)\und\tens\cdots\und\tens A(R)$
or a quotient if it by determinants etc.  Here $\und\tens$ has to be specified
but it is natural to take here the braided-tensor product algebra as introduced
in the theory of braided groups. We have explained in \cite{Ma:inf} that if one
takes the braiding $\Psi(\vect_1\tens\vect_2)=
\vect_2\tens\vect_1 R$ corresponding to $A(R)$ as a right $A(R)$-comodule by
right-comultiplication, then one has from the definition of braided tensor
product algebras as in (\ref{btens}) that the algebra $A(R)^{\und\tens \infty}$
has generators $\vect(i)$ and relations
\eqn{kac-moody}{\vect_1(i)\vect_2(j)=\cases{\vect_2(j)\vect_1(i) R^{-1}_{21}
&if\ $i<j$\cr R_{12}^{-1}\vect_2(j)\vect_1(i)R_{12}& if\ $i=j$\cr
\vect_2(j)\vect_1(i)R_{12}& if\ $i>j$}.}
This is essentially the wave-function algebra algebra in \cite{AFS:hid} as we
have remarked in \cite{Ma:inf}.

One would still like the point of view whereby such braided tensor products are
derived automatically rather than put in by hand. Such a point of view is the
following, at least for the Hecke case such as corresponding to $G=GL_n$. We
think of $LG$ directly as a wave-function or field on $S^1$. From this point of
view it is natural to model it as a $n\times\infty$ (or $n\times nN$)
rectangular quantum matrix. This consists of consecutive $n\times n$ blocks,
each corresponding to one site on $S^1$. For the rows then we naturally take
the $GL_n$ $R$-matrix. For the columns we take the $GL_{\infty}$ R-matrix
$R_\infty$, or more precisely a $GL_{nN}$ R-matrix to be finite.

\begin{propos} The rectangular quantum matrix algebra $A(R:R_\infty)$ can be
identified by cutting into $n\times n$ blocks with the algebra
$A(R)^{\und\tens^\infty}$ in (\ref{kac-moody}).
\end{propos}
\proof This is an application of the general theory of glueing of quantum block
matrices introduced by the author and M. Markl in \cite{MaMar:glu}. There we
show that the $GL_{nN}$ R-matrix can be  built up as $R\oplus_q\cdots\oplus_q
R$ where $\oplus_q$ is a certain associative glueing operation among Hecke
R-matrices (which we also introduce). On the other hand, we show in \cite[Sec.
4]{MaMar:glu} that $A(R_1:R_2\oplus_q R_3)=A(R_1:R_2)\und\tens_{R_1}
A(R_1:R_3)$ where the braided tensor product is with braiding given by $R_1$.
Iterating this formalism in our present example gives the result.  \endproof

Note that the discussion supposes that a limit $N\to \infty$ can also be taken
in some way. The present remarks are purely algebraic and we do not address
this point here. On the other hand (at the $M_q(n)$ level) one can now add the
wave-functions pointwise with braid statistics. One can also differentiate,
both with respect to the field and in the group indices alone by using the
techniques above. These are some of the ingredients used when formulating
classical $\sigma$-models. In 2D-quantum gravity one also has exchange algebras
which can be described as braided tensor product algebras\cite{Ma:inf} and
hence as rectangular quantum matrices in the same way.
Thus the general scheme of using rectangular quantum matrices to systematically
make q-deformed lattice approximations in field theory, as explained at the end
of Section~4, seems to apply fairly generally. Moreover, the results in
\cite{MaMar:glu} apply generally to provide a local description as quantum
blocks with braided tensor product statistics between them. This indicates an
interesting direction for further work.

%\bibliographystyle{unsrt}
%\bibliography{biblio}

\end{document}